\documentclass[letterpaper, 10 pt, conference]{ieeeconf}  

\IEEEoverridecommandlockouts                             

\usepackage{graphicx} 
\usepackage[font=footnotesize]{caption} 
\usepackage{amsmath} 
\usepackage{multirow} 

\title{\LARGE \bf
Optimizing the Level of Challenge in Stroke Rehabilitation using Iterative Learning Control: a Simulation
}

\author{Sandra-Carina Noble$^{1}$, Tomas Ward$^{2}$ and John V. Ringwood$^{1}$
\thanks{*This work is supported by the Irish Research Council under project ID GOIPG/2020/692.}
\thanks{$^{1}$Sandra-Carina Noble and John V. Ringwood are with the Department of Electronic Engineering, Maynooth University, Maynooth, Ireland
        {\tt\small sandracarina.noble.2017@mumail.ie, john.ringwood@mu.ie}}%
\thanks{$^{2}$Tomas Ward is with the School of Computing, Dublin City University, Dublin, Ireland
        {\tt\small tomas.ward@dcu.ie}}%
}

\begin{document}

\onecolumn
$\copyright$ 2021 IEEE.  Personal use of this material is permitted.  Permission from IEEE must be obtained for all other uses, in any current or future media, including reprinting/republishing this material for advertising or promotional purposes, creating new collective works, for resale or redistribution to servers or lists, or reuse of any copyrighted component of this work in other works.

\newpage
\twocolumn

\maketitle
\thispagestyle{empty}
\pagestyle{empty}

\begin{abstract}
    The level of challenge in stroke rehabilitation has to be carefully chosen to keep the patient engaged and motivated while not frustrating them. This paper presents a simulation where this level of challenge is automatically optimized using iterative learning control. An iterative learning controller provides a simulated stroke patient with a target task that the patient then learns to execute. Based on the error between the target task and the execution, the controller adjusts the difficulty of the target task for the next trial. The patient is simulated by a nonlinear autoregressive network with exogenous inputs to mimic their sensorimotor system and a second-order model to approximate their elbow joint dynamics. The results of the simulations show that the rehabilitation approach proposed in this paper results in more difficult tasks and a smoother difficulty progression as compared to a rehabilitation approach where the difficulty of the target task is updated according to a threshold.
\end{abstract}

\section{Introduction}
It is understood that a stroke patient's motivation can affect the outcome of their rehabilitation \cite{Rapoliene2018}. Physical therapists and other rehabilitation professionals have reported that they try to motivate a patient by controlling the task difficulty and therefore helping the patient gain confidence in their abilities \cite{Oyake2020}. If a task is too easy, the patient might become bored, whereas if it is too difficult, they might become frustrated \cite{Oyake2020}. 

In robotic rehabilitation, the level of challenge is often maintained by the assist-as-needed approach, where the robot only applies as much assistance as the patient needs to successfully complete a task, thus avoiding slacking and reducing the level of assistance as the patient regains their abilities \cite{Mounis2019}. Once virtual reality or serious games are involved in rehabilitation, the actual task is commonly adapted based on a rule-based system or using machine learning \cite{Zahabi2020}.

Using a robot or exoskeleton is not always feasible and increasing the task difficulty in fixed steps, as is often done in rule-based approaches, might not lead to an optimal level of challenge. That is why this paper presents how the level of challenge in stroke rehabilitation can be optimized using iterative learning control (ILC), where the update step is adjusted according to the patient's ability, based on a simulation. ILC is typically applied to repetitive systems, such as industrial robots, that repeat the same task over several trials. It uses the tracking error and input of past trials to compute the input for the next trial, thus eventually eliminating the tracking error \cite{Ahn2007}. In the past, ILC has been applied to stroke rehabilitation to control the level of functional electrical stimulation applied to the patient's affected limb \cite{Freeman2012a}, or to control the assistance provided by a robot \cite{Marchal-Crespo2009}. In \cite{Choi2008}, a heuristic approach that resembles ILC has been used to control the time a patient has to complete a task. 

\begin{figure*}[htpb]
    \centering
    \includegraphics[width=\textwidth]{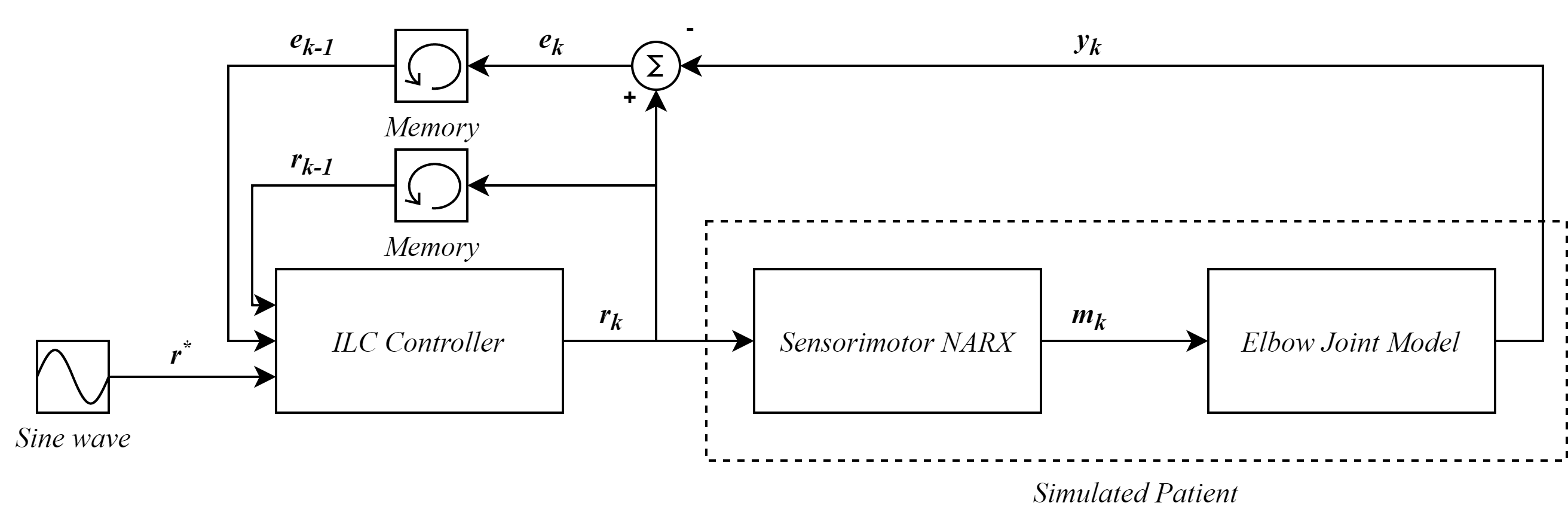} 
    \caption{System overview}
    \label{fig:overview}
\end{figure*}

Fig. \ref{fig:overview} shows an overview of the simulation, which consists of two parts; the iterative learning controller and the simulated stroke patient. The iterative learning controller provides the simulated patient with a target task for their affected upper limb, which the patient then attempts to learn. This task is a cyclical movement of the forearm on a planar surface, so that it is represented by a sine wave. The amplitude of the sine wave is trial-varying with a maximum amplitude of \(0.2\) radians, whereas its angular frequency remains constant at \(\frac{2}{3} \pi\) radians per second. The controller evaluates the position error between the target task and the patient's movement and provides a new target task based on this error, the previous target task and the exercise goal to achieve in the rehabilitation session. In this simulation, the patient's sensorimotor system is simulated by a nonlinear autoregressive network with exogenous inputs (NARX) and their elbow joint dynamics are approximated by a second-order model. Visual perception models have not been included in the simulation as they are believed to have minor effects on the overall system since human motion, which is simulated in this study, is well within the visible spatio-temporal range of humans \cite{VandenBrandenLambrecht1996} and perception is not the focus of this study. 

\section{Simulated Stroke Patient}
The simulated stroke patient processes the target task given by the ILC controller in the sensorimotor system, which is simulated by a NARX network. This system then provides a motor command to the second-order elbow joint model, which executes the learned movement.

\subsection{Sensorimotor System}
Two `sensorimotor' NARX neural networks have been implemented. They differ only in the number of hidden layers and the number of nodes per layer as the overall number of nodes is consistent in both networks.

The inputs to the NARX networks are the target motor command, which is derived from the target task, and past values of its output. The output time-series is the motor command, which is passed to the elbow joint model. $NARX_1$ has one hidden layer with 7 nodes and $NARX_2$ has two hidden layers, where the first layer consists of 4 nodes and the second layer consists of 3 nodes. The hidden layer(s) use the sigmoid activation function and the output layer uses a linear activation function. Both networks are trained with Bayesian regularization backpropagation for 100 epochs and in a closed-loop fashion. 



Network lesioning, where a node or edge is removed from a graph, has been used in the past to study the effects of stroke \cite{Aerts2016,Straathof2019}. In \cite{Rismanchi2018}, cortical lesions were simulated by removing nodes in an artificial neural network.

As the number of nodes in the input and output layers of the NARX are fixed, network lesioning can only be applied to the hidden layer(s). Table \ref{table:nodes} shows the number of nodes that were removed to simulate a stroke. The nodes were removed after the networks were fully trained on the ultimate target task.

\begin{table}[h]
\caption{Number of removed nodes in each hidden layer in simulated stroke}
\label{table:nodes}
\begin{center}
\begin{tabular}{|c | c | c |}
\hline
& \multicolumn{2}{c|}{Nodes removed in} \\
 &  1st layer & 2nd layer \\
\hline
$NARX1$ & 3 & N/A\\
\hline
$NARX_2$ & 2 & 1 \\
\hline
\end{tabular}
\end{center}
\end{table}

\subsection{Elbow Joint Model}
The output of the `sensorimotor' NARX, in the form of a motor command, is passed to the elbow joint model. While Hill's muscle model \cite{Hill1938} is commonly preferred over a second-order model for human joint dynamics, due to its increased accuracy and detail on muscle-level \cite{Winters1987}, a second-order model is sufficient for this simulation as the focus of this work is on the improvement of motor learning, rather than the study of human movement. The model of the elbow joint dynamics used in this simulation is based on \cite{Abe2003} but modified to exclude any gravitational effects as the task is a horizontal movement on a planar surface, with friction ignored. Therefore, the model equation is
\begin{equation} \label{eq:elbModel}
    \tau(t) = J\ddot{\theta } + B\dot{\theta} + K \theta
\end{equation}
where \(\tau(t)\) is the motor torque command, \(\theta\) is the joint angle and \(J\), \(B\) and \(K\) are the inertia, (linearised \cite{Zatsiorsky1997}) viscosity and stiffness of the joint, respectively. Table \ref{table:jbk} gives the values used in the model. These are the mean values of the elbow-apparatus system identified in \cite{Abe2003}. Since a general elbow joint model is sufficient for the simulations presented in this paper, the effects of the apparatus on the model parameters have been ignored.

\begin{table}[h]
\caption{Parameters for elbow joint model}
\label{table:jbk}
\begin{center}
\begin{tabular}{|c|c|}
\hline
Inertia $J$ (kg m$^2$) & 0.144 $\pm$ 0.014 \\
\hline
Viscosity $B$ (Nms/rad) & 0.22 $\pm$ 0.10 \\
\hline
Stiffness $K$ (Nm/rad) & 4.96 $\pm$ 1.16 \\
\hline
\end{tabular}
\end{center}
\end{table}

\section{Iterative Learning Controller}
After each trial, the ILC controller uses the instantaneous position error between the target task and actual movement to update the target task. The update law is
\begin{equation} \label{eq:simILC}
    r_k = r_{k-1} + (\alpha r^*)(1-\beta ||e_{k-1}||_2)
\end{equation}
where \(k\) denotes the trial number, \(r\) is the target task and \(r^*\) refers to the ultimate target task to achieve in the rehabilitation session, \(||e||_2\) is the $\ell_2$-norm of the error, and \(\alpha\) and \(\beta\) are adjustable parameters.

\(\alpha\) controls the maximum update step that happens only when there is zero error. It is expressed as a percentage of the ultimate target task. Table \ref{table:alpha} shows how \(\alpha\) affects the update step in simulations with a pre-trained $NARX_1$, when \(\beta\) is kept at \(1.0\) and the simulated patient is healthy, i.e. no nodes in the network have been removed. The $\ell_2$-norm of the error of the previous trial, $||e_{k-1}||_2$, is shown in brackets. It should be noted that there is no previous error in the first trial as there is no previous data, it is always nominally set to \(\frac{1}{\beta}\).

\begin{table}[h]
\vspace{0.2cm}
\caption{Amplitude of target task (in rad) for different \(\alpha\) values}
\label{table:alpha}
\begin{center}
\begin{tabular}{|c|c|c|c|c|c|}
\hline
\(\alpha\) & Trial 1 & Trial 2 & Trial 3 & Trial 4 & Trial 5\\
\hline
\multirow{2}{*}{0.2} & 0.040 & 0.071  & 0.095 & 0.114 & 0.128\\
 &  (1.00) & (0.222) & (0.398) & (0.533) & (0.638) \\
 \hline
\multirow{2}{*}{0.3} & 0.040 & 0.087 & 0.118 & 0.138 & 0.152 \\
& (1.00) & (0.222) & (0.486) & (0.659) & (0.773) \\
\hline
\end{tabular}
\end{center}
\end{table}
    
\(\beta\) controls how much effect the error has on the update step. The term \(\beta ||e_{k-1}||_2\) is capped at unity so that, if the error is large, the target task does not shrink to zero. Table \ref{table:beta} shows how different \(\beta\) values affect the update of the target task, when \(\alpha\) is kept at \(0.2\) and the simulated patient is healthy. Again, the $\ell_2$-norm of the previous error is shown in brackets.

\begin{table}[h]
\caption{Amplitude of target task (in rad) for different \(\beta\) values}
\label{table:beta}
\begin{center}
\begin{tabular}{|c|c|c|c|c|c|}
\hline
\(\beta\) & Trial 1 & Trial 2 & Trial 3 & Trial 4 & Trial 5\\
\hline
\multirow{2}{*}{0.5} & 0.040 & 0.076 & 0.107 & 0.135 & 0.160\\
& (2.00) & (0.222) & (0.423) & (0.600) & (0.757) \\ 
\hline
\multirow{2}{*}{1.5} & 0.040 & 0.067 & 0.084 & 0.096 & 0.104 \\
& (0.67) & (0.222) & (0.374) & (0.472) & (0.539)\\
\hline
\end{tabular}
\end{center}
\end{table}

\section{Results}
To evaluate the use of ILC to optimize the level of challenge in stroke rehabilitation, the ILC approach described in this paper has been compared to a rule-based approach \cite{Zahabi2020}, where the target task is updated only if the $\ell_2$-norm of the error in the previous trial is below a certain threshold, in this case 0.7. Therefore, the update law for the rule-based approach is
\begin{equation} \label{eq:non-ILC}
    r_k =
    \begin{cases}
        r_{k-1} + (\alpha r^*) \textit{ , if } ||e_{k-1}||_2 \leq 0.7 \\
        r_{k-1} \textit{ , otherwise}
    \end{cases}
\end{equation}
where the terms are the same as in (\ref{eq:simILC}).

All simulations were run for 20 trials with \(\alpha = 0.2\) and \(\beta = 1.0\). The target task was a 30 second sine wave sampled at 100 Hertz with varying amplitude and a frequency of \(\frac{2}{3} \pi\) radians per second. 

\begin{figure}[htpb]
        \includegraphics[width=\linewidth]{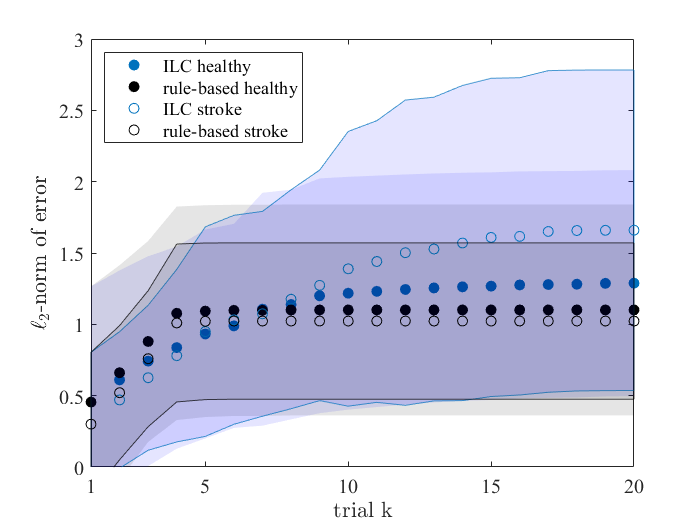} 
    \captionof{figure}{Mean $\ell_2$-norm of error over trials for $NARX_1$}
    \label{fig:narx1_mean}
\end{figure}

The simulations were repeated for 100 different sets of initial weights and biases of both NARX networks, respectively. Fig. \ref{fig:narx1_mean} illustrates the mean $\ell_2$-norm of the error between the target task and the actual movement over trials for $NARX_1$. The shaded areas without border indicate the standard deviation for a simulated healthy patient and the shaded areas with border show the standard deviation for a simulated stroke patient.

\begin{figure}[htpb]
        \includegraphics[width=\linewidth]{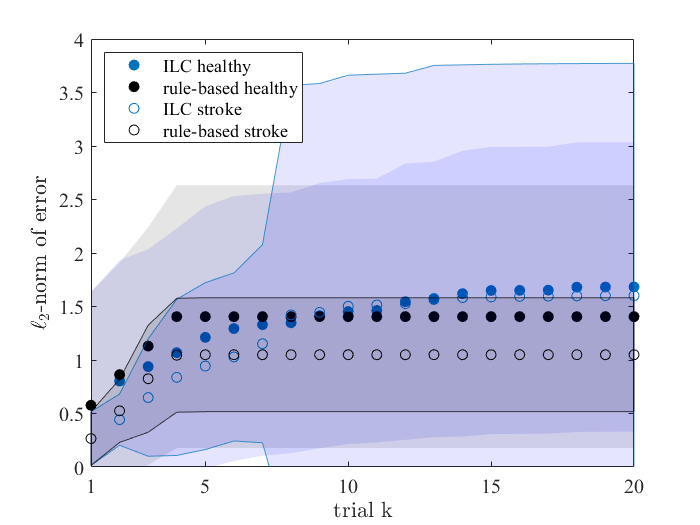} 
    \captionof{figure}{Mean $\ell_2$-norm of error over trials for $NARX_2$}
    \label{fig:narx2_mean}
\end{figure}

\begin{figure}[htpb]
        \includegraphics[width=\linewidth]{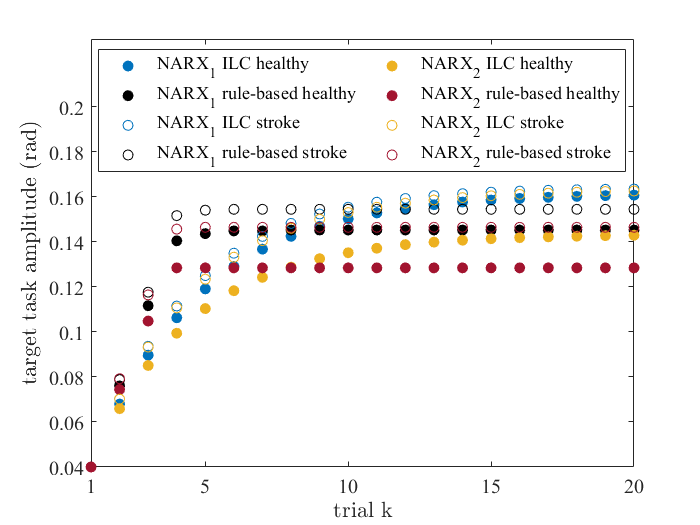} 
    \captionof{figure}{Mean target task amplitude over trials for all scenarios}
    \label{fig:amp}
\end{figure}

The mean $\ell_2$-norm of the error that was achieved with $NARX_2$ can be seen in Fig. \ref{fig:narx2_mean} and the mean target task amplitude that was used in each trial is shown in Fig. \ref{fig:amp} for all scenarios. The average standard deviations for the scenarios in Fig. \ref{fig:amp} can be found in Table \ref{table:std}.

\begin{table}[htpb]
\vspace{0.2cm}
\caption{Standard deviation corresponding to mean target task amplitude (in rad), averaged over trials, for each scenario}
\label{table:std}
\begin{center}
\begin{tabular}{|c|c|c|c|c|}
\hline
& \multicolumn{2}{c|}{Healthy patient} & \multicolumn{2}{c|}{Stroke patient} \\
 & ILC & rule-based & ILC & rule-based\\
\hline
$NARX_1$ & 0.032 & 0.033 & 0.021 & 0.020\\
\hline
$NARX_2$ & 0.039 & 0.040 & 0.022 & 0.024 \\ 
\hline
\end{tabular}
\end{center}
\vspace{-0.5cm}
\end{table}

Figs. \ref{fig:narx1_mean}, \ref{fig:narx2_mean} and \ref{fig:amp} illustrate that, on average, the simulated stroke patient performs better, i.e. lower mean errors and therefore higher target task amplitudes, than the healthy patient in almost all cases, which does not represent reality.

\begin{figure}[htpb]
        \includegraphics[width=\linewidth]{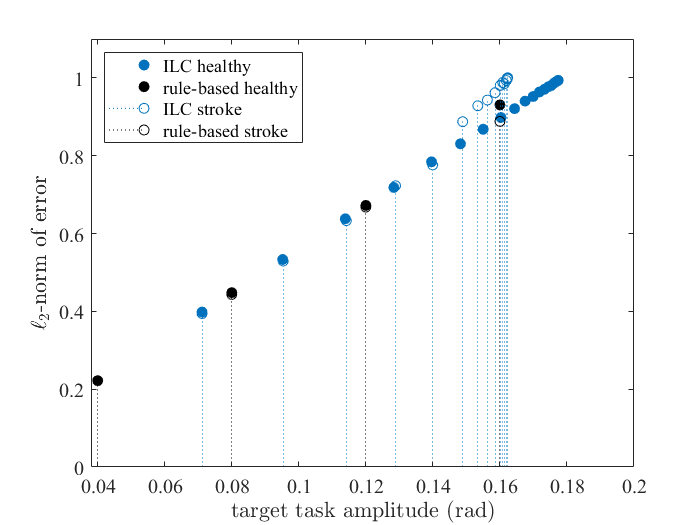} 
    \captionof{figure}{$\ell_2$-norm of error over target task amplitude for $NARX_1$}
    \label{fig:narx1_example}
\end{figure}

\begin{figure}[htpb]
        \includegraphics[width=\linewidth]{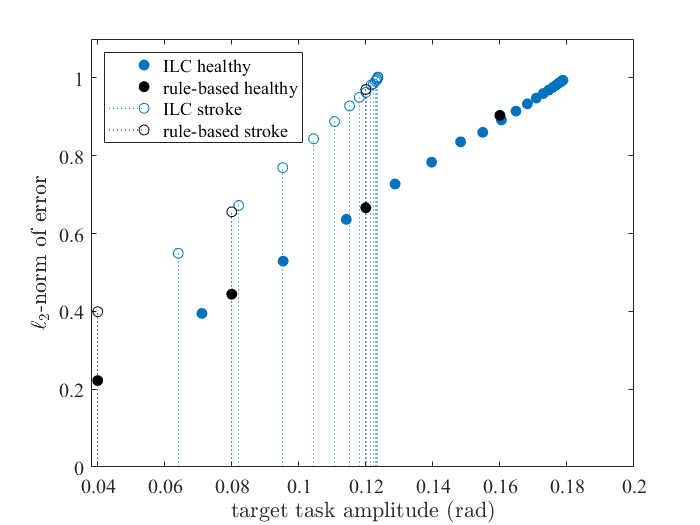} 
    \captionof{figure}{$\ell_2$-norm of error over target task amplitude for $NARX_2$}
    \label{fig:narx2_example}
\end{figure}

Fig. \ref{fig:narx1_example} shows a specific example of the $\ell_2$-norm of the error over target task amplitudes for $NARX_1$. Where the same amplitude was used in several trials, the mean of the errors is shown. An example of simulation results for $NARX_2$ is illustrated in Fig. \ref{fig:narx2_example}.

\vspace{-0.75cm}

The errors produced by the healthy and the stroke patients for both the ILC and rule-based approaches follow a similar pattern, however, the ILC algorithm results in smaller update steps and a greater final task amplitude as it keeps updating the reference until the error exceeds $\frac{1}{\beta}$, or in this case unity, whereas the rule-based approach stops updating once the error exceeds 0.7.

\section{Conclusions}
This paper presents an ILC-based approach to optimally adapt the level of difficulty of a target task in stroke rehabilitation. While the simulation results show that NARX networks in combination with network lesioning are not ideal for simulating stroke as the lesioned networks resulted in better performances than the healthy networks on average, it has been shown that the ILC-based approach leads to more difficult final tasks and smaller update steps, especially as the error approaches unity, compared to a rule-based approach. These results indicate that a stroke patient may become more frustrated with the latter approach as the task difficulty is increased in large steps, even if the patient is already starting to struggle with the task. In contrast, the ILC-based rehabilitation approach reduces the update step as the error increases, i.e. as the patient begins to struggle with the task. This is believed to improve patient motivation and confidence. Human experiments need to be conducted to test this hypothesis.

\bibliographystyle{./IEEEtran} 
\bibliography{./IEEEabrv,./IEEEBib}

\begin{thebibliography}{10}
\providecommand{\url}[1]{#1}
\csname url@rmstyle\endcsname
\providecommand{\newblock}{\relax}
\providecommand{\bibinfo}[2]{#2}
\providecommand\BIBentrySTDinterwordspacing{\spaceskip=0pt\relax}
\providecommand\BIBentryALTinterwordstretchfactor{4}
\providecommand\BIBentryALTinterwordspacing{\spaceskip=\fontdimen2\font plus
\BIBentryALTinterwordstretchfactor\fontdimen3\font minus
  \fontdimen4\font\relax}
\providecommand\BIBforeignlanguage[2]{{%
\expandafter\ifx\csname l@#1\endcsname\relax
\typeout{** WARNING: IEEEtran.bst: No hyphenation pattern has been}%
\typeout{** loaded for the language `#1'. Using the pattern for}%
\typeout{** the default language instead.}%
\else
\language=\csname l@#1\endcsname
\fi
#2}}

\bibitem{Rapoliene2018}
J.~Rapolienė, E.~Endzelytė, I.~Jasevičienė, and R.~Savickas, ``Stroke
  patients motivation influence on the effectiveness of occupational therapy,''
  \emph{Rehabil. Res. Pract.}, vol. 2018, July 2018.

\bibitem{Oyake2020}
K.~Oyake, M.~Suzuki, Y.~Otaka, and S.~Tanaka, ``Motivational strategies for
  stroke rehabilitation: A descriptive cross-sectional study,'' \emph{Front.
  Neurol.}, vol.~11, June 2020.

\bibitem{Mounis2019}
S.~Y. Mounis and N.~Z. Azlan, ``{Assist-as-needed control strategies for upper
  limb rehabilitation therapy: A review},'' \emph{Jurnal Mekanikal}, vol.~42,
  pp. 57--74, June 2019.

\bibitem{Zahabi2020}
M.~Zahabi and A.~M. {Abdul Razak}, ``{Adaptive virtual reality-based training:
  a systematic literature review and framework},'' \emph{Virtual Real.},
  vol.~24, pp. 725--752, Dec. 2020.

\bibitem{Ahn2007}
H.~S. Ahn, Y.~Q. Chen, and K.~L. Moore, ``{Iterative learning control: Brief
  survey and categorization},'' \emph{{IEEE} Trans. Syst., Man, Cybern. {C}},
  vol.~37, no.~6, pp. 1099--1121, Nov. 2007.

\bibitem{Freeman2012a}
C.~T. Freeman, E.~Rogers, A.~M. Hughes, J.~H. Burridge, and K.~L. Meadmore,
  ``{Iterative learning control in health care: Electrical stimulation and
  robotic-assisted upper-limb stroke rehabilitation},'' \emph{{IEEE} Control
  Syst. Mag.}, vol.~32, no.~1, pp. 18--43, Feb. 2012.

\bibitem{Marchal-Crespo2009}
L.~Marchal-Crespo and D.~J. Reinkensmeyer, ``{Review of control strategies for
  robotic movement training after neurologic injury},'' \emph{J. Neuroeng.
  Rehabil.}, vol.~6, no.~20, June 2009.

\bibitem{Choi2008}
Y.~Choi, F.~Qi, J.~Gordon, and N.~Schweighofer, ``{Performance-based adaptive
  schedules enhance motor learning},'' \emph{J. Mot. Behav.}, vol.~40, no.~4,
  pp. 273--280, 2008.

\bibitem{VandenBrandenLambrecht1996}
C.~J. {van den Branden Lambrecht}, ``{A working spatio-temporal model of the
  human visual system for image restoration and quality assessment
  applications},'' \emph{1996 IEEE International Conference on Acoustics,
  Speech, and Signal Processing Conference Proceedings}, vol.~4, pp.
  2291--2294, 1996.

\bibitem{Aerts2016}
H.~Aerts, W.~Fias, K.~Caeyenberghs, and D.~Marinazzo, ``{Brain networks under
  attack: Robustness properties and the impact of lesions},'' \emph{Brain},
  vol. 139, no.~12, pp. 3063--3083, Dec. 2016.

\bibitem{Straathof2019}
M.~Straathof, M.~R. Sinke, A.~van~der Toorn, P.~L. Weerheim, W.~M. Otte, and
  R.~M. Dijkhuizen, ``{Differences in structural and functional networks
  between young adult and aged rat brains before and after stroke lesion
  simulations},'' \emph{Neurobiol. Dis.}, vol. 126, pp. 23--35, June 2019.

\bibitem{Rismanchi2018}
M.~Rismanchi, ``{The inhibitory effect of functional lesions on eloquent brain
  areas: from research bench to operating bed},'' \emph{Int. J. Neurosci.},
  vol. 128, no.~11, pp. 1022--1029, Nov. 2018.

\bibitem{Hill1938}
A.~V. Hill, ``{The heat of shortening and the dynamic constants of muscle},''
  \emph{Proc. R. Soc. Lond. B}, vol. 126, no. 843, pp. 136--195, Oct. 1938.

\bibitem{Winters1987}
J.~M. Winters and L.~Stark, ``{Muscle models: What is gained and what is lost
  by varying model complexity},'' \emph{Biol. Cybern.}, vol.~55, pp. 403--420,
  1987.

\bibitem{Abe2003}
M.~O. Abe and N.~Yamada, ``{Modulation of elbow joint stiffness in a vertical
  plane during cyclic movement at lower or higher frequencies than natural
  frequency},'' \emph{Exp. Brain Res.}, vol. 153, no.~3, pp. 394--399, Dec.
  2003.

\bibitem{Zatsiorsky1997}
V.~M. Zatsiorsky, ``{On muscle and joint viscosity},'' \emph{Motor Control},
  vol.~1, no.~4, pp. 299--309, Oct. 1997.

\end{thebibliography}

\end{document}